\documentclass{aastex}

\usepackage{amsfonts}
\usepackage{amssymb}
\usepackage{amsmath}
\usepackage{amsbsy}
\usepackage{graphicx}
\usepackage{natbib}
\usepackage[english]{babel}

\newcommand{\vect}[1]{{\bf #1}}
\newcommand{\bnabla}{\boldsymbol{\nabla}}

\newcommand{\md}[1]{\mathcal #1}

\newcommand{\sech}{{\rm sech\,}}

\newcommand{\ey}{\,\boldsymbol{{\rm e}}_y}

\newcommand{\ud}{\mathrm{d}}

\newcommand{\dt}[1]{\frac{\partial #1}{\partial t}}

\newcommand{\lunit}{\ensuremath{\,\mbox{km}}}
\newcommand{\vunit}{\ensuremath{\,\mbox{km} \, \mbox{s}^{-1}}}

\newcommand{\nunit}{\ensuremath{\,\mbox{cm}^{-3}}}
\newcommand{\tunit}{\ensuremath{\,\mbox{s}}}
\newcommand{\efluxunit}{\ensuremath{\,\mbox{erg}/\mbox{cm}^2\,\mbox{s}}}

\newcommand{\pfluxunit}{\ensuremath{\,\mbox{particle}/\mbox{cm}^2\,\mbox{s}}}

\newcommand{\ba}{\begin{eqnarray}}
\newcommand{\ea}{\end{eqnarray}}

\newcommand{\bas}{\begin{eqnarray*}}
\newcommand{\eas}{\end{eqnarray*}}

\newcommand{\be}{\begin{equation}}
\newcommand{\ee}{\end{equation}}

\newcommand{\bes}{\begin{equation*}}
\newcommand{\ees}{\end{equation*}}

\newcommand{\bfs}{\begin{figure*}[t!]}
\newcommand{\efs}{\end{figure*}}
\newcommand{\bfd}{\begin{figure}[tbh]}
\newcommand{\bft}{\begin{figure}[t!]}
\newcommand{\bfh}{\begin{figure}[h!]}
\newcommand{\bfb}{\begin{figure}[b!]}
\newcommand{\ef}{\end{figure}}

\newcommand{\bd}{\begin{displaymath}}
\newcommand{\ed}{\end{displaymath}}

\shorttitle{Non-steady reconnection in the Sun}

\begin{document}

\title{Spontaneous non-steady magnetic reconnection within the solar environment} 

\author{Lapo Bettarini \altaffilmark{1} and Giovanni Lapenta \altaffilmark{1}}
\email{lapo.bettarini@wis.kuleuven.be}

\affil{Centrum voor Plasma-Astrofysica, Departement Wiskunde, Katholieke Universiteit Leuven, Celestijnenlaan 200B, 3001 Leuven, Belgium}


\begin{abstract}
This work presents a 2.5-dimensional simulation study of the instability of current-sheets located in a medium with a strong density variation along the current layer. The initial force-free configuration is observed to undergo a two-stage evolution consisting of an abrupt regime transition from a slow to a fast reconnection process leading the system to a final chaotic configuration. Yet, the onset of the fast phase is not determined by the presence of any anomalous enhancement in plasma's local resistivity, but rather is the result of a new mechanism discovered in~\citet{lapenta08} and captured only when sufficient resolution is present. Hence, the effects of the global resistivity, the global viscosity and the plasma beta on the overall dynamics are considered. This mechanism allowing the transition from slow to fast reconnection provides a simple but effective model of several processes taking place within the solar atmosphere from the high chromosphere up to the low corona. In fact, the understanding of a spontaneous transition to a self-feeding fast reconnection regime as well as its macroscopic evolution is the first and fundamental step to produce realistic models of all those phenomena requiring fast (and high power) triggering events.
\end{abstract}

\keywords{MHD - Sun: activity - Sun: chromosphere - Sun: magnetic field}

\section{Introduction} \label{sec1}

Magnetic reconnection is a fundamental mechanism playing a major role both in the dynamics of astrophysical objects as well as in the evolution of several structures in the solar environment, in planetary magnetospheres and in the heliosphere. Here, indeed, a broad range of phenomena undergoes a multi-stage evolution determined by a rapid transition from a slow to a fast, and often bursty, dynamics which presents typical magnetic reconnection marks as the change in the magnetic field-line connectivity, the formation of high-velocity plasma jets, plasma heating, and particle accelerations~\citep{biskamp93}.

Magnetic reconnection is believed to be the key process underlying the evolution of a number of processes that take place in the solar environment. In particular, it may be the responsible for the release of magnetic flux and plasma in the corona and in the interstellar medium during solar explosive phenomena. In general, a catastrophic loss of mechanical equilibrium of the underlying coronal magnetic structure is required and magnetic reconnection plays a role either in triggering the initiating eruption or in the aftermath of the event by characterizing the subsequent evolution and related structures like post CME's current-sheets~\citep{forbes00,lin04}. In the case of a flare evolution, the free energy of the solar magnetic field (exceeding the energy of the solar atmosphere potential field) is released. This excess energy is connected with currents flowing in the corona from the photosphere throughout the chromosphere and the transition region, the flare process being the process of rapid variation of these currents. 

From \textit{Yohkoh} X-ray telescope observations~\citep{shibata96}, we have evidences consistent with the reconnection model of solar flares~\citep{tsuneta96}. In particular, the onboard soft X-ray telescope revealed many jet-like features in which the out-flowing plasma is associated with a change in the magnetic field topology of the underlying X-ray emitting structure and with the rise of an X-type neutral point location as the reconnection process goes on. Recently, Hinode observations have allowed a more detailed picture of X-jet formation and dynamics~\citep{cirtain07} and they are suggesting that the heating of the solar chromosphere and corona may be related to small-scale ubiquitous reconnection~\citep{shibata07}. \citet{karpen98} provided a detailed scenario for the formation of current-sheets and their dissipation through reconnection during a chromospheric erupting event. They pointed out that magnetic reconnection occurs in a bursty and intermittent mode whose unknown physical mechanism is in contrast with usual Sweet-Parker (SP, hereafter) and Petschek (PE, hereafter) models of reconnection and it is essential to explain the time evolution of flaring events. Their 2.5-dimensional numerical simulations of magnetic reconnection in the lower solar atmosphere showed also the formation of two-sided outflows whose dynamics and acceleration should be affected by the vertical density variation from the chromosphere to corona's upper layers. These bi-directional jets have been observed in the form of simultaneous blue and red Doppler shifts and they are characteristic structures present in most of the UV explosive solar events~\citep[and references therein]{aschwanden06book}.

Large-scale chromosphere/corona models are carried out within MHD. Yet, it is widely accepted that in order to obtain a fast reconnection regime in MHD, it is necessary to have a localized enhancement of the plasma resistivity~\citep{hautz87,scholer89,kliem00}. In fact, it is believed that the description of a plasma as a resistive fluid is not sufficient to obtain the fast reconnection rate expected by the models, for instance, of the above mentioned structures. On the other hand, the microscopic mechanisms leading to the formation of the invoked anomalous resistivity depend on several factors and have not been identified conclusively. New insights on this topic have been recently given by \citet{baty06} who claim to have maintained a PE-like reconnection process in two-dimensional time-dependent MHD simulations without the use of a significantly high localized resistivity. However, the onset of this process depends still on a non-uniform, even though small, resistivity. Furthermore, \citet{nitta07} shows that it is possible to have in MHD a self-similar reconnection process with a continuous transition from a slow reconnection regime to a fast one by increasing the magnetic Reynolds number. He considers a new series of solutions of two-dimensional MHD equations on a wide range of the magnetic Reynolds number, that is from about $10$ to $2100$. The reconnecting structures change according to the regime the system undergoes, passing from an X-point configuration, to an X-O-X point, that is two X-points with a magnetic islands placed in between. Finally, for the slowest regime at the highest Reynolds number the magnetic islands collapse forming a long current-sheet with a Y profile. 

On the other hand, \citet{lapenta08} observes the spontaneous development of a fast reconnection mechanism on macroscopic scales. A two-dimensional current-sheet evolves through two different stages, that is a first slow very elongated SP layer is followed by a rapid transition to a fast chaotic reconnection process, and the system does not need anomalous resistivity nor driven flows to undergo this evolution. This MHD bursty process leading to a chaotic final state is due to the destabilization of the SP layer via the tearing instability as predicted in previous theoretical studies~\citep{furth63,bulanov79,loureiro07}. The resulting chaotic reconnection phase is fast and has the fundamental property of being independent both from the Lundquist number as well as from the dynamic Reynolds number. In order to observe such dynamics, the horizontal size of our numerical box must be large with respect to the thickness of the forming SP layer and also a high Lundquist number has to be considered. This allows the SP layer to expand till the instability driving the system to a chaotic reconnection regime sets in. During this phase, an outflow pattern determining a circulation loop feeding-back reconnection regions is observed. The net results obtained by the development of such process are that (a) it is much faster, developing on scales of the order of the Alfv\'en time, and (b) the areas of reconnection become distributed chaotically over a macroscopic region. Furthermore, recent three-dimensional MHD numerical analyses~\citep{kowal09} test and confirm the investigation on the effects of turbulence on MHD magnetic reconnection~\citep{lazarian99,kowal09}: in the presence of weakly stochastic magnetic fields, the speed of the process is independent both of Ohmic and anomalous resistivity.

The above described picture does not include the effects of a density gradient either parallel or orthogonal to the two-dimensional current-sheet. In compressible conditions, the presence of a density variation may influence the onset and the evolution of reconnecting field lines and the resulting jet formation and acceleration. By introducing a stratification due to a gravitational field, the reconnection process evolves towards a traditional two-dimensional pattern passing through an initial stage whose features and duration depend on the imposed stratification~\citep{galsgaard02}. In general, a lower reconnection rate is found and it decreases according to a faster magnetic field expansion which produces a more complex reconnection dynamics. 

The present work has as primary goal to study the spontaneous fast chaotic reconnection process in the presence of a strong density gradient modeling the solar atmosphere from the high chromosphere up to the low corona. There, the rapid density variation is well approximated with a constant plasma pressure~\citep{syrovatskii81} where the temperature gradient balances the density variation. Hence, the gravitational field can be neglected and a force-free configuration of the magnetic field is usually assumed. This reconnection process via a chaotic evolution is a promising candidate to explain the cross-scale process linking the small-scale reconnection dynamics to the triggering and ejection of large-scale explosive phenomena, such as such as solar flares and coronal mass ejections (CME hereafter)~\citep{tajimashibatabook}. We move from the two-dimensional MHD model of a current-sheet in the high chromosphere/ low corona provided by~\citet{yokoyama01} and we determine the conditions and the features of a spontaneous transition from a slow reconnection regime to a fast dynamics leading to a chaotic macroscopic state of the reconnection site.
 
In the following sections, we introduce the numerical setup, the initial conditions, the perturbations, and the parameters for our analysis (Sect.~\ref{sec2}). In Sect~\ref{sec3} and \ref{sec4}, we present the results of the simulations for the several cases. We observe the development of a two-stage process with the onset of a fast reconnection. The dynamics proceeds with the fracture of the initial reconnection site in several structures determining a final turbulent state. Moreover, the density jump from the chromosphere to the corona is critical in giving an asymmetric development of the global structure, determining an oriented path for the overall flow and current-density evolution. We observe also an overall final accelerating jet outward to the upper layers of the corona. In Sect.~\ref{sec5}, we summarize the results drawing our conclusions and pointing to fundamental open questions and future directions.

\section{Numerical settings} \label{sec2}

We define a longitudinal or stream-wise direction ($z$), oriented away from the sun, wherein open boundary conditions are set at $z = 0$ and $L_z$, and a transverse or cross-stream direction ($x$) along which all the equilibrium quantities vary and where we set reflecting boundary conditions. Hence, we solve numerically the viscous-resistive compressible $2.5$D MHD equations. In dimensionless form we have the following set of equations
\ba
\label{ro}
\dt{\rho} & = & -\bnabla \cdot \rho \vect{v} \\
\label{vel}
\rho \dt{\vect{v}} & = & - \rho \left(\vect{v} \cdot \bnabla\right) \vect{v} -\bnabla p-\left[\bnabla \frac{|\vect{B}|^2}{2}- \left(\vect{B} \cdot \bnabla\right) \vect{B}\right] + \frac{1}{\md{R}_\nu} \Delta \vect{v} \\
\label{magn}
\dt{B} & = & \bnabla \times \left(\vect{v} \times \vect{B}\right) + \frac{1}{\md{S}} \Delta \vect{B} \\
\label{energy}
\rho \dt{\md{I}} & = & -p \bnabla \cdot \vect{v} + \frac{\left(\bnabla \cdot \vect{v} \right)^2}{\md{R}_\lambda} + \frac{\Pi \cdot \Pi}{\md{R}_\nu} + \frac{\vect{J} \cdot \vect{J}}{\md{S}} \, ,
\ea
where $\rho$ is the mass density, $\vect{B}$ is the magnetic field, $\vect{v}$ is the fluid velocity, $\md{I}$ is the specific internal energy, $\vect{J}$ is the current density, $p$ is the fluid pressure, and $\Pi$ is the symmetric rate-of-strain tensor defined as
\be
\Pi = \frac{1}{2} \left[\bnabla \vect{v} + \bnabla \vect{v}^T\right] \, .
\ee
In Eq.~\eqref{vel}-\eqref{energy}, $\md{R}_\lambda$ and $\md{R}_\nu$ are the Reynolds numbers measuring respectively the global kinematic and dynamic viscosity in the numerical box: in our simulations we always consider $\md{R}_\lambda = \md{R}_\nu$ and so we refer only to the dynamic Reynolds number $\md{R}_\nu$. The quantity $\md{S}$ is the Lundquist number measuring the global explicit resistivity we set within the domain. We use an ideal equation of state with polytropic index $\gamma = 5/3$ such that $p = \rho \, (\gamma-1) \, \md{I}$. The equations are normalized according to reasonable values for the high chromosphere/low solar corona: the number density $\rho/m \approx 10^9 \nunit$; the length scale $L \approx 1000-2000 \lunit$, where $L  = \mbox{current-sheet width} = L_x/20 = L_z/80$; the Alfv\'en velocity $v_A \approx 400-600 \vunit$ (approximately three times the sound speed). So, we have a time scale of $t_A = L/v_A \approx 1.5-5 \tunit$.

We use the three-dimensional code, FLIP MHD~\citep{brackbill91}. We consider $600$ (in $x$) $\times$ $960$ (in $z$) lagrangian markers arrayed initially in a $3 \times 3$ uniform formation in each of the $200 \times 320$ cells of our numerical grid. This has the following sizes
 \be
 \begin{array}{ll}
 x \in [-10,10] & \quad \mbox{with} \, \ud x = 0.1 \\
 z \in [0,80] & \quad \mbox{with} \, \ud z = 0.25 \, .
 \end{array}
 \ee 
As shown in Fig.~\ref{fig1}, our system consists in a current-sheet determined by a force-free field configuration given by the following equations
\ba
\label{b0z}
B_{0z}(x) & = & \tanh x \\
\label{b0y}
B_{0y}(x) & = & -\sech x \, ,
\ea
which implies the current-sheet width is equal to the normalization length $L$ and it is resolved by $30$ Lagrangian markers. As already pointed out in the introduction, our computational domain has transversal sizes large enough to allow a complete development of the system dynamics. The density is modeled as a step function
\be
\label{rhostep}
\rho = 
\left\{
\begin{array}{ll}
\md{C} & \qquad \mbox{if} \qquad z < z_0  = L_z/8\\
& \\
1 & \qquad \mbox{if} \qquad z \ge z_0
\end{array}
\right .
\ee
or as a hyperbolic tangent function
\be
\label{rhotanh}
\rho = \frac{1+\md{C}}{2}+\frac{\left(1-\md{C}\right)}{2} \tanh\left(\frac{z-z_0}{\alpha}\right) \, ,
\ee
where $\md{C}$ is the ratio between the density in the region $0 < z < L_z/8$ (modeling the high chromosphere) and that one in $L_z/8 < 0 < L_z$ (the low corona) and we choose it to range from $10^2$ up to $10^5$ as considered in~\citet{yokoyama01}. In Eq.~\eqref{rhostep}, $z_0$ is the position of the discontinuity resembling the transition region, whereas in Eq.~\eqref{rhotanh} it represents the transition region's center; the parameter $\alpha$ allows us to assume different thicknesses of the transition region.

Reconnection is set in by a magnetic field perturbation defined by the out-of-plane component of the magnetic vector potential
\be
\label{pertb}
\vect{B}_1 = \bnabla \times \left( A_{1y} \ey \right)
\ee
with
\ba
\label{pertay}
A_{1y} & = & \varepsilon \exp\left[-\kappa_x (x-x_p)^2-\kappa_z (z-z_p)^2\right] \cos\kappa_x (x-x_p) \cos\kappa_z (z-z_p)
\ea
providing a GEM-like perturbation~\citep{birn01} but localized in $(x, z) = (x_p, z_p)$ by the exponential factor. We referred to the model and simulations of~\citet{yokoyama01}, but with the intention to study the possibility of a natural-evolving slow-fast reconnection transition in the low corona with the development of a final turbulent state. So, after having performed a few test simulations, we set the initial perturbation amplitude ($\varepsilon$) to $0.5$ and the perturbation wave-numbers in the $x$ ($\kappa_x$) and $z$ ($\kappa_z$) direction to $\pi/L_x$ and $\pi/4 L_z$ respectively. In general, we consider that the perturbation is located exactly at the center of the $x-$ and $z-$axis, that is in the low corona: $(x_p, z_p) = (0, L_z/2)$. Other cases with different $z_p$ ($L_z/8, L_z/4, 0)$ but always $x_p = 0$ have been considered and performed. 

\section{Simulation parameters} \label{sec3}

\subsection{Viscosity and resistivity} \label{sec3.1}

Fig.~\ref{fig2} shows the reconnected flux as a function of the simulation time for several different values of the Lundquist and dynamic Reynolds number, both ranging from $10^2$ to $\infty$, even though the actual upper limit is given by the intrinsic numerical effects. The plasma $\beta$ is set for these simulation to $0.2$ (see \S~\ref{sec3.2}). As we use a non-selfconsistent perturbation that is not an eigenstate of the system, we have a small transient time of about a decade of time steps. Afterwards, the system displays the same behavior, except for very low $\md{S}$ or $\md{R}_\nu$: it undergoes a two-stage evolution consisting in a slow reconnection phase followed by a rapid increase of the reconnected flux which corresponds to a fast reconnecting regime. The viscosity and resistivity affect the dynamics of the system in a different way. Viscosity is effective only when it is set to high values ($\md{R}_\nu = 10^2$). Resistivity has the main role in determining both the behavior of the system in the first slow phase and in switching it to a fast regime. If $\md{S}$ is too low, as described by the dotted line in the above mentioned figure, the system remains in a steady-state reconnection process where the corresponding width of our current-sheet is constant in time until the end of the simulation. By increasing $\md{S}$, the slow phase shortens and we observe the triggering of a fast regime whose starting point and dynamical features appear to be independent of the value of resistivity. If we increase $\md{S}$ further until the numerical dissipation dominates ($\md{S} \rightarrow \infty$), the two-stage evolution continues (as long as viscosity has a negligible value).

As long as the reconnection process remains in the first slow reconnection regime, the behavior of the system for different values of the resistivity is well-described by the SP theory. In fact, the scaling of the reconnection rate, $\md{M}$, prescribed by SP theory is that $\md{M} = \md{S}^{-1/2} = {\eta/(\md{L} v_A)}$, where $\eta$ is the plasma diffusivity, $\md{L}$ the longitudinal size of the diffusion region and $v_A$ is the upstream Alfv\'en speed. In Fig.~\ref{fig3}, we show the details of the slow regime (as in the small frame of Fig.~\ref{fig2} but) for the curves corresponding to $\md{S} = 10^4$ (solid line), $\md{S} = 10^3$ (dotted line) and $\md{S} = 10^2$ (dashed line). The observed reconnection rates are consistent with the expected behavior, even though our basic field and perturbation settings does not allow to measure correctly $\md{L}$ and to consider it fixed for all the simulations.

In the subsequent discussion, the focus will be on $\md{S} = \md{R}_\nu = 10^4$.

\subsection{The plasma beta} \label{sec3.2}

As previously pointed out, we consider the simplified assumption, yet reasonable for the solar environment, that a temperature gradient is provided in order to balance the density variation. Hence, we have an overall uniform and constant initial plasma pressure
\be
\label{p0}
p_0 = \beta \frac{|B_0|^2}{2} = \frac{\beta}{2} \, .
\ee
In Fig.~\ref{fig4}, we show the reconnected flux as a function of the simulation time for different values of the beta parameter ranging from $0.2$ to $2$. As the Alfv\`en velocity is fixed, the change of $\beta$ implies we are varying the compressibility of our system. In the first slow and diffusive SP reconnection regime, the difference in the compressibility does not influence the evolution at all. In fact, the dynamics depends on the global system's resistivity/viscosity that is fixed by $\md{R}_\nu = \md{S} = 10^4$. Yet, as soon as we observe the triggering of the second faster reconnection phase, which corresponds to system's structure evolving on smaller and smaller length-scales (see below), it is evident that for higher compressibility, the slope of the reconnected flux becomes steeper and hence the corresponding reconnection rate becomes faster. So, the main features of the fast-reconnection phase depend critically on the compressive effects in the system and no longer on the global resistivity/viscosity. For realistic simulation below, we consider the accepted value $\beta = 0.2$ to place the current-sheet approximately in the high chromosphere/low corona~\citep{yokoyama01,gary01}.

\section{Results and discussion} \label{sec4}

The evolution of a current-sheet in the presence of an orthogonal strong density variation is characterized by a two-stage evolution whose main features are described in Fig.~\ref{fig5}. The contour of the out-of-plane component of the current density ($J_y$) is shown superposed on the same component of the magnetic vector potential (solid black lines) at four different instants. At about $t \approx 18-20$ (top panel), the system is evolving according to the SP theory (see \S~\ref{sec3.1}) and the slow reconnection regime proceeds with the elongation and thinning of the current sheet in the low density region (second panel of Fig.~\ref{fig5}, $t \approx 36$). As described in the previous work by~\citet{bulanov79} and in recently ones~\citep{loureiro07,skender09}, the system reaches a configuration such that the ratio between the length and the width of the current sheet makes it tearing-unstable. This is evident also in Fig.~\ref{fig6} where diffusion region's aspect ratio is shown as a function of the simulation time: at about $t \approx 40$ a maximum is reached, the instability process rapidly evolves and the initially formed SP diffusion region starts to destabilize. Due to the disruption of the main current sheet, several structures form each in turn determining a reconnection site (third panel from the top of Fig.~\ref{fig5}, $t \approx 52$): those drive the initial configuration to smaller and smaller scales (bottom panel). During all the process, the presence of the strong density gradient produces an asymmetric development determining an oriented path for the overall flow and current density evolution. As the two-stage reconnection proceeds, we observe the slow upward movement of the initial diffusion region while it is breaking up into several parts. The islands that form start to move forward and backward pushed by the reconnection jets and a chaotic configuration is more and more evident both in the most intense current-density region still in fragmentation as well as in the macro-structure emitted towards the upper layers of the corona and that resembles the turbulent reconnection as proposed by~\citet{tajimashibatabook} or~\citet{lazarian99}. Furthermore, due to the presence of the density wall the buildup of magnetic flux of opposite polarity is taking place and post-disruption arcades are forming in the low corona. As the process goes on, converging motions towards the polarity inversion line lead to the shrinking of the current-sheet at the base of these growing magnetic structures. This situation probably leads to a release of magnetic energy in the solar atmosphere via a new reconnection process.

The onset of a macroscopic circulation pattern linking all the small reconnection sites sustains the strong increase of the reconnection rate as seen in Fig.~\ref{fig2} (case with $\md{S} = 10^4$). The net effect is an overall acceleration of the system that is dragging forward all the structures and therefore expelling plasma in the upper corona. In Fig.~\ref{fig7}, the contour of the out-of-plane magnetic vector potential at $t \approx 96$ is shown with the velocity field superposed and represented by properly scaled arrows: we can observe the well-developed circulation pattern during the multiple reconnection process and the effective acceleration reconnection jet. The motions driving the post-disruption arcades rising in the low corona are also evident.

\subsection{Reconnection fluxes} \label{sec4.1}

The energy transfers provide information on the structures moving down towards the high chromosphere or up towards the upper layers of the solar corona. We consider the flux of the Poynting vector, the enthalpy flux and the kinetic flux, respectively defined as
\ba
\label{poynting}
\vect{S} & = & \vect{E} \times \vect{B} \\
\label{enthalpy}
\vect{H} & = & 	(\md{I} + p) \, \vect{v} \\
\label{kinetic}
\vect{K} & = & \frac{1}{2} \rho \, |\vect{v}|^2 \, \vect{v}
\ea
through a surface extended from $x = -5$ to $x = +5$ and located at $z = 20$ (downward flux) and $z = 80$, the higher boundary (upward flux). In Fig.~\ref{fig8}, it is possible to observe these profiles as a function of time through the two-stage process: the top panel describes the upward flux, whereas the bottom panel describes the downward flux. The mass flux through the same surfaces is shown as reference. We observe that the fast process determines a sensitive rise in the upward mass flux associated with an increase of the kinetic and enthalpy fluxes: the multiple chaotically-moving plasmoids accelerate upward and the plasma is characterized by an increase of the compressional heating effects. The chaotic fast stage shows a initial increase of all fluxes and a weak maximum is reached. Afterwards, a real impulsive event occurs and it has a duration of about $\Delta t \approx 10$ that is $15 \div 50 \tunit$ according to the normalization values provided in \S~\ref{sec2}. Furthermore, we obtain the following physical values for the energy fluxes associated with the impulse
\ba
\rho \, v_z & \approx & 4 \quad \Rightarrow 0.8 \div 2.4 \cdot 10^{16} \pfluxunit\\
H_z & \approx & 3 \quad \Rightarrow 0.6 \div 6.6 \cdot 10^8 \efluxunit \\
K_z & \approx & 1 \quad \Rightarrow 0.2 \div 2.2 \cdot 10^8 \efluxunit
\ea
which show a reasonable agreement with observations of solar explosive phenomena~\citep{tsuneta96,shibata96}. The Poynting up-flux increases during most of the fast process and it reaches a constant value of about $S_z \sim 1.2$, that is $0.2 \div 2.6 \cdot 10^8 \efluxunit$, but it drops out in the impulsive phase. So, the rapid emission of a highly structured accelerating mass towards the upper layers of the corona does not come with a corresponding impulsive release of magnetic energy that is instead converted into the other forms of energy. With regards to the downward-fluxes, only small variations are observed. This behavior is due to the presence of the density gradient that, as already discussed, allows the pile-up of magnetic flux with the formation of post-disruption arcades. In Fig.~\ref{fig9}, it is possible to observe the contour of the specific internal energy at $t \sim 96$ with respect its initial value, $\md{I}/\md{I}(t=0)$. It is evident the turbulent configuration of the most dense structures whose internal energy rapidly increased during the evolution. In particular, we note the high values obtained on the post-disruption arcades in the low corona.

\subsection{Effect of perturbation location} \label{sec4.2}

As already pointed out at the end of \S~\ref{sec2}, we performed simulations with different values of $z_p$, but no substantial qualitative differences are present as a function of the position of the initial perturbation, at least in the cases considered. For instance, if perturbation is located on the transition region ($x_p = 0$ and $z_p = L_z/8$) modeled as a discontinuity between the chromosphere and the corona, we observe the same current-sheet dynamics as described in the previous sections, even though two clear differences can be observed in Fig.~\ref{fig10}, where the out-of-plane current density (in color) and magnetic vector potential (solid black lines) are shown for this case at $t \approx 96$: 
\begin{itemize}
\item the density step does not allow the current-sheet to develop into the high chromosphere at all and this determines that  the whole structure is elongated towards upper coronal layers. The accelerating jet still presents a chaotic internal structure, but it has a wider aperture with respect the previous cases.
\item the motion of the overall structure towards the upper coronal layer allows again the formation of post-disruption arcades that develop slower than they do in the previous cases. By the end of the simulation, we are not able to determine any visible sign of a starting reconnection process at the base of the corona ($L_z \approx 10$).
\end{itemize}
The upward fluxes in this case do not change with respect the previous ones, while the downward fluxes are affected by the presence of the density step determining the accumulation of magnetic flux that encompasses the surface through which the flux is computed.

In the case the density is modeled by an hyperbolic tangent, the same dynamics is observed as long as a steep profile is considered: the smoother the slope of the density function is, the less sensitive the accumulation of magnetic flux, the rise of the main reconnection site and the overall acceleration of system is. Finally, with a very low density variation, we recover the results obtained by~\citet{lapenta08}, even though the presence of open boundary conditions instead of periodic settings alters the detail of the evolution.

\section{Conclusions} \label{sec5}

We presented a 2.5-dimensional model of the current-sheet evolution located in the solar atmosphere, in particular across the region where the chromosphere and the corona meet. The transition region is modeled as a density and temperature discontinuity or with an hyperbolic tangent profile. Previous studies of such systems, most of which focusing on triggering events and dynamics of solar flares~\citep{yokoyama01}, on filaments~\citep{lin08} or on the evolution and observations of current-sheets forming in the aftermath of coronal mass ejections~\citep{poletto04,bemporad06}, always demanded the presence of locally enhanced plasma resistivity to explain the rapid increase of the reconnection rate to allow the fast evolution underlying explosive phenomena in the solar atmosphere. Yet, in the present paper we demonstrate a natural two-stage evolution of a current-sheet with a spontaneous transition from a low to a fast reconnection regime without the need to include any anomalous effect and in the presence of a strong density gradient.

We analyzed such mechanism by means of high resolution simulations. We observed how the step density jump in the transition region produces an asymmetric development of the structure, determining an oriented path for the overall flow and current density evolution. Despite the simplifications of our model, the global dynamics reported resembles the tearing evolution of solar structures both just before and immediately after observed and/or modeled explosive phenomena. Several observational clues can be mentioned to support the scenario described in the present paper. For example during solar flares, hard X-ray and radio emissions present many properties ascribed to the presence of small-scale, fragmented, ÒburstyÓ magnetic reconnections~\citep{aschwanden02}. Post-CME current-sheets are observed to have a much larger thickness than what MHD theory prescribes~\citep{bemporad08}. Chaotic magnetic reconnection in a current-sheet could explain this observed behavior. 

In the future, an ``expanding box'' setting~\citep{grappin93,rappazzo05} of our simulations may be considered in order to follow the evolution of the current-sheet in a simpler Cartesian geometry with spherical effects included. This would allow us also to properly study the formation and evolution of outward and downward flows as seen in H$\alpha$ observations of highly dynamic filaments and surges in active regions~\citep{lin08}. Furthermore, the analysis of the frequency distribution of time scales can possibly discriminate between a fractal current-sheet evolution or turbulent reconnection, since the former process is scale-free and it generally produces power law distributions, while turbulent processes are controlled by incoherent random processes and hence they generally give exponential distributions~\citep{aschwanden06book}. Further extension to three-dimensions is needed to fully address realistic configurations.
  

\acknowledgments
The research leading to these results has received funding from the European Commission's Seventh Framework Programme (FP7/2007-2013) under the grant agreement n$^\circ$ 218816 (SOTERIA project, {\tt \verb+www.soteria-space.eu+}). The simulations shown were conducted using processors on the VIC cluster of the Vlaams Supercomputer Centrum at K.U. Leuven (Belgium). 


\clearpage

\bfh
\begin{center}
\includegraphics[width=0.3\textwidth]{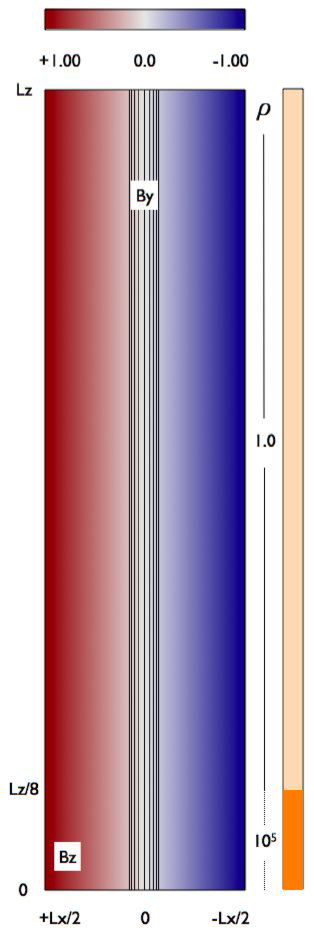}
\caption{(color online) Schematic reconstruction of our basic system: it is possible to visualize the initial stream-wise component of the magnetic field (in blue and red) and the density step at $L_z/8$.}
\label{fig1}
\end{center}
\ef

\bfh
\begin{center}
\includegraphics[width=0.6\textwidth]{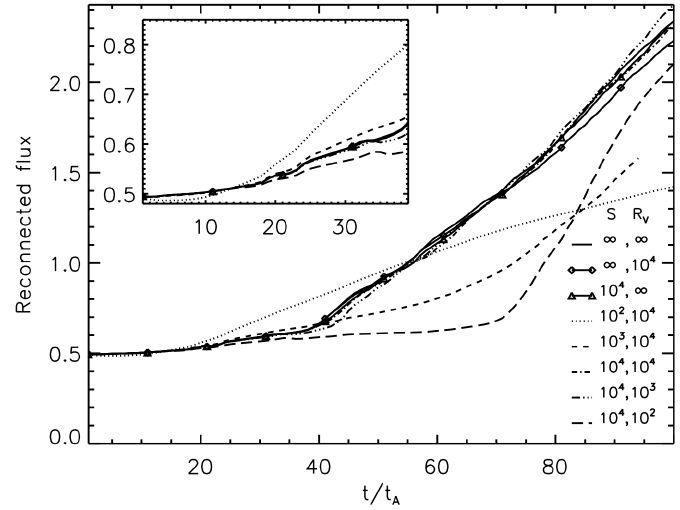}
\caption{Big plot: reconnected flux as a function of the simulation time for different values of the Lundquist number, $\md{S}$, and the dynamic Reynolds number, $\md{R}_\nu$. In all cases but $(\md{S}, \md{R}_\nu) = (10^2, 10^4)$ (dotted line), we can observe an abrupt transition from a slow reconnection process to a fast regime. Small plot: detailed view of the slow SP regime.}
\label{fig2}
\end{center}
\ef

\bfh
\begin{center}
\includegraphics[width=0.6\textwidth]{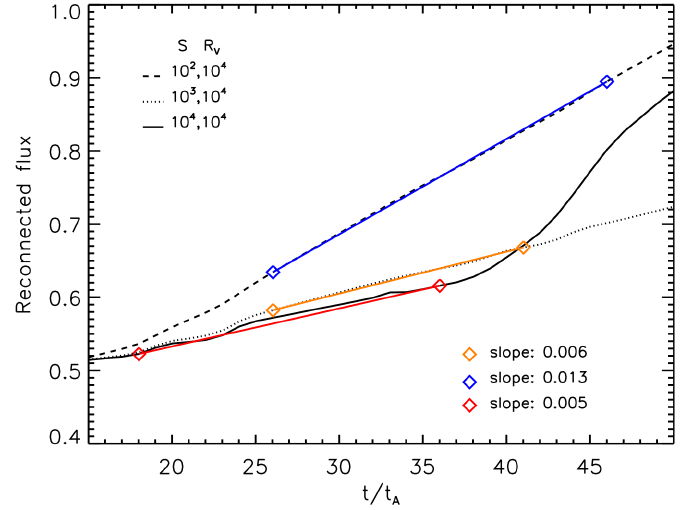}
\caption{(color online) Reconnected flux as a function of the simulation time but $t < 50$ (slow SP regime) for three cases among those shown in Fig.~\ref{fig2}. Reconnected rates are calculated and shown.}
\label{fig3}
\end{center}
\ef

\bfh
\begin{center}
\includegraphics[width=0.6\textwidth]{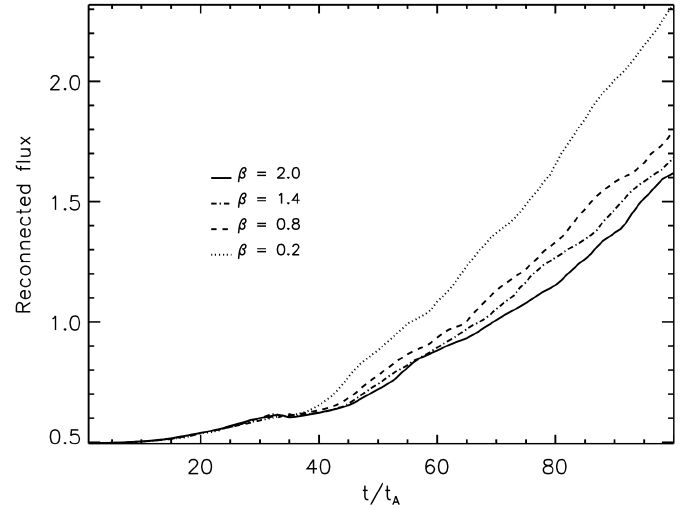}
\caption{Reconnected flux as a function of the simulation time for different values of the beta parameter with the Lundquist number and the dynamic Reynolds number set to $\md{S} = \md{R}_\nu = 10^4$.}
\label{fig4}
\end{center}
\ef

\bfh
\begin{center}
\includegraphics[width=0.8\textwidth]{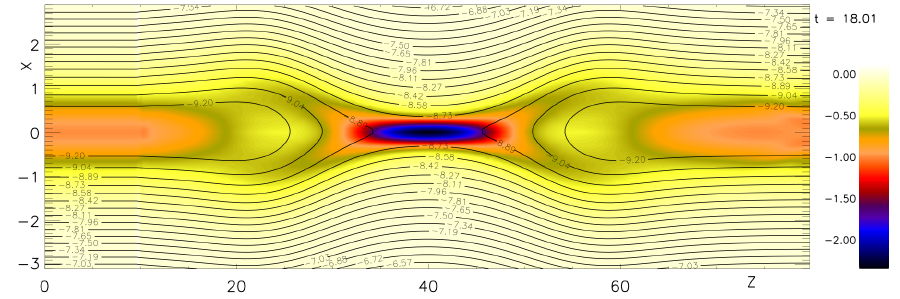}
\includegraphics[width=0.8\textwidth]{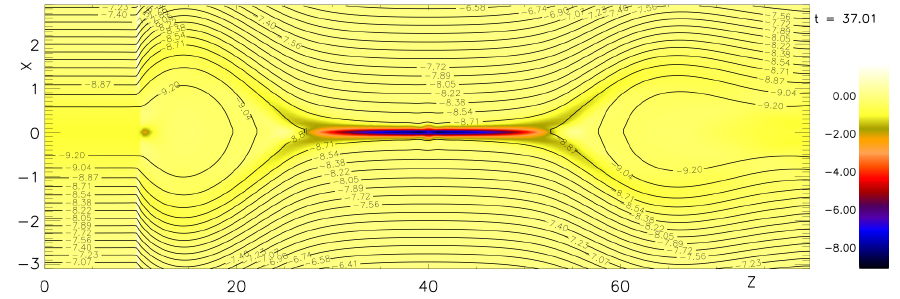}
\includegraphics[width=0.8\textwidth]{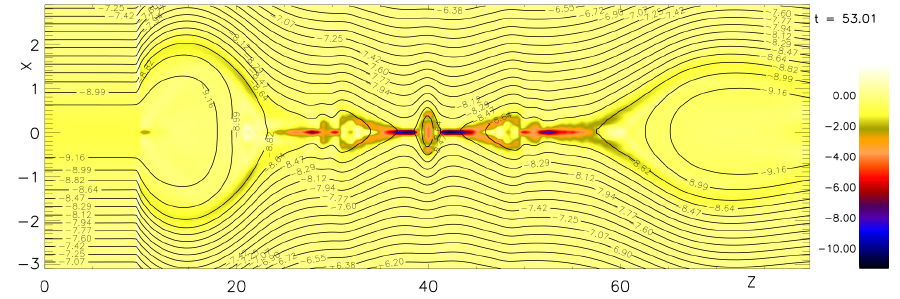}
\includegraphics[width=0.8\textwidth]{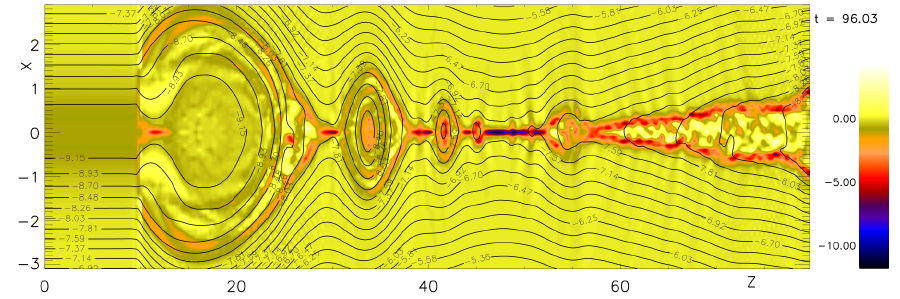}
\caption{(color online) Contour of the out-of-plane current density, $J_y$, (in color) superposed on the same component of the vector potential (solid black lines) at four different instants (from top to bottom): $t = 17, 36, 53, 96$. Here, the density is modeled as a discontinuity with the step located at $z = L_z/8 = 10$. For a better visualization, only the central part of the numerical box is shown in the images.}
\label{fig5}
\end{center}
\ef

\bfh
\begin{center}
\includegraphics[width=0.6\textwidth]{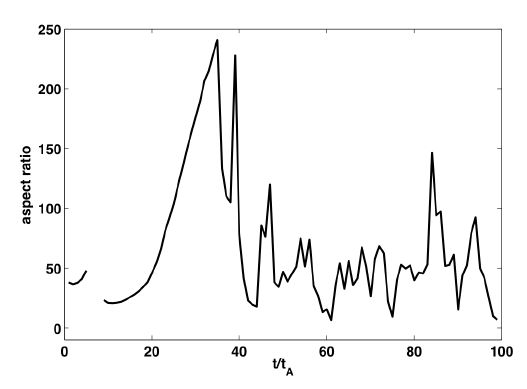}
\caption{Diffusion region's aspect ratio (length/width) as a function of the simulation time.}
\label{fig6}
\end{center}
\ef

\bfh
\begin{center}
\includegraphics[width=0.8\textwidth]{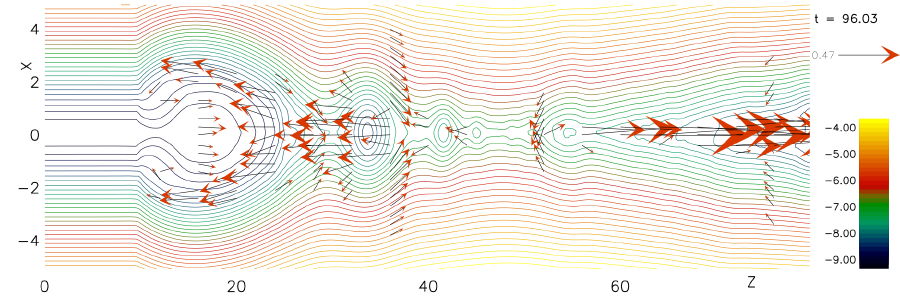}
\caption{(color online) Contour of the out-of-plane magnetic vector potential at $t \approx 96$ (solid colored lines). The velocity field is also shown and it is represented by properly scaled arrows. For a better visualization, only the central part of the numerical box is shown in the image.}
\label{fig7}
\end{center}
\ef

\bfh
\begin{center}
\includegraphics[width=0.5\textwidth]{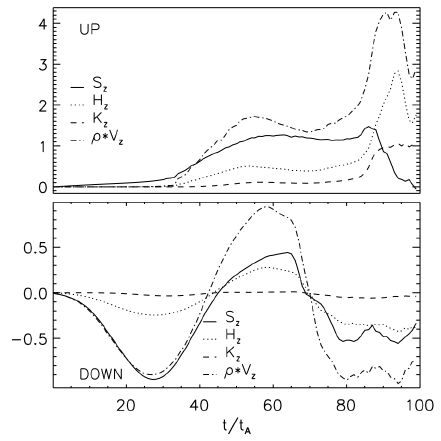}
\caption{Top panel: energy and mass fluxes through a surface extended from $x = -5$ to $x = +5$ and located at the higher boundary, $z = 80 = L_z$ (upward fluxes). Bottom panel: same as the top panel, but the surface for the computation of the fluxes is located at $z = 20$ (downward fluxes).}
\label{fig8}
\end{center}
\ef

\bfh
\begin{center}
\includegraphics[width=0.8\textwidth]{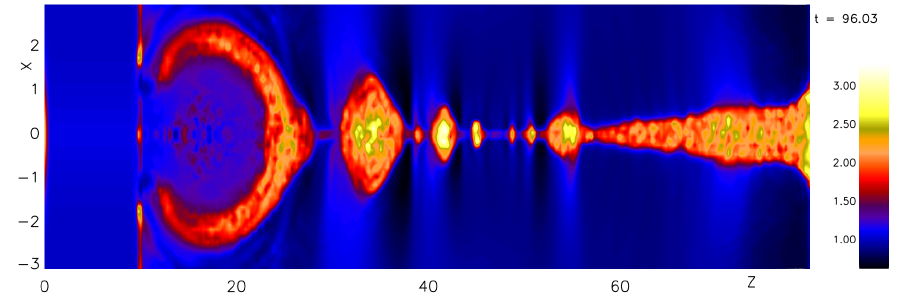}
\caption{(color online) Contour of the specific internal energy with respect its initial value, $\md{I}/\md{I}(t=0)$, at $t = 96$. For a better visualization, only the central part of the numerical box is shown.}
\label{fig9}
\end{center}
\ef

\bfh
\begin{center}
\includegraphics[width=0.8\textwidth]{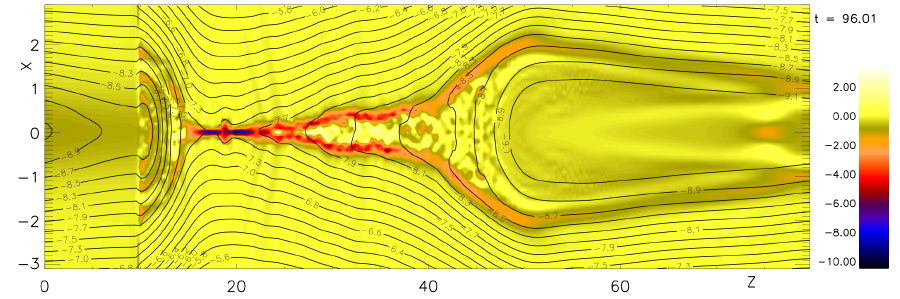}
\caption{(color online) Contour of the out-of-plane current density, $J_y$, (in color) superposed on the same component of the vector potential (solid black lines) at $t = 96$. Here, the density is modeled as a discontinuity with the step located at $z = L_z/8 = 10$ and the initial perturbation is located at $x_p = 0, z_p = L_z/8$. For a better visualization, only the central part of the numerical box is shown.}
\label{fig10}
\end{center}
\ef


\begin{thebibliography}{35}
\expandafter\ifx\csname natexlab\endcsname\relax\def\natexlab#1{#1}\fi

\bibitem[{Aschwanden(2002)}]{aschwanden02}
Aschwanden, M.~J. 2002, Space Sci. Rev., 101, 1

\bibitem[{Aschwanden(2006)}]{aschwanden06book}
---. 2006, Physics of the solar corona: an introduction with problems and
  solutions (Springer Science \& Business)

\bibitem[{Baty {et~al.}(2006)Baty, Priest, \& Forbes}]{baty06}
Baty, H., Priest, E.~R., \& Forbes, T.~G. 2006, Phys. Plasmas, 13, 022312

\bibitem[{Bemporad(2008)}]{bemporad08}
Bemporad, A. 2008, ApJ, 689, 572

\bibitem[{Bemporad {et~al.}(2006)Bemporad, Poletto, Suess, Ko, Schwadron,
  Elliott, \& Raymond}]{bemporad06}
Bemporad, A., Poletto, G., Suess, S.~T., Ko, Y.-K., Schwadron, N.~A., Elliott,
  H.~A., \& Raymond, J.~C. 2006, ApJ, 638, 1110

\bibitem[{Birn {et~al.}(2001)Birn, Drake, Shay, Rogers, Denton, Hesse,
  Kuznetsova, Ma, Bhattacharjee, Otto, \& Pritchett}]{birn01}
Birn, J., {et~al.} 2001, J. Geophys. Res., 106, 3715

\bibitem[{Biskamp(1993)}]{biskamp93}
Biskamp, D. 1993, Nonlinear magnetohydrodynamics, Cambridge Monograph on Plasma
  Physics (Cambridge, England: Cambridge University Press)

\bibitem[{Brackbill(1991)}]{brackbill91}
Brackbill, J.~U. 1991, J. Comp. Phys., 96, 163

\bibitem[{Bulanov {et~al.}(1979)Bulanov, Sakai, \& Syrovatskii}]{bulanov79}
Bulanov, S.~V., Sakai, D.-I., \& Syrovatskii, S.~I. 1979, Soviet Journal of
  Plasma Physics, 5, 280

\bibitem[{Cirtain {et~al.}(2007)Cirtain, Golub, Lundquist, van Ballegooijen,
  Savcheva, Shimojo, DeLuca, Tsuneta, Sakao, Reeves, Weber, Kano, Narukage, \&
  Shibasaki}]{cirtain07}
Cirtain, J.~W., {et~al.} 2007, Science, 318, 1580

\bibitem[{Forbes \& Lin(2000)}]{forbes00}
Forbes, T., \& Lin, J. 2000, J. Atmos. Solar-Terr. Phys., 62, 1499

\bibitem[{{Furth} {et~al.}(1963){Furth}, {Killeen}, \& {Rosenbluth}}]{furth63}
{Furth}, H.~P., {Killeen}, J., \& {Rosenbluth}, M.~N. 1963, Phys. Fluids, 6,
  459

\bibitem[{Galsgaard \& Roussev(2002)}]{galsgaard02}
Galsgaard, K., \& Roussev, I. 2002, A\&A, 383, 685

\bibitem[{Gary(2001)}]{gary01}
Gary, G.~A. 2001, Sol. Phys., 203, 71

\bibitem[{Grappin {et~al.}(1993)Grappin, Velli, \& Mangeney}]{grappin93}
Grappin, R., Velli, M., \& Mangeney, A. 1993, Phys. Rev. Lett., 70, 2190

\bibitem[{Hautz \& Scholer(1987)}]{hautz87}
Hautz, R., \& Scholer, M. 1987, Geophys. Res. Lett., 14, 969

\bibitem[{Karpen {et~al.}(1998)Karpen, Antiochos, DeVore, \& Golub}]{karpen98}
Karpen, J.~T., Antiochos, S.~K., DeVore, C.~R., \& Golub, L. 1998, ApJ, 495,
  491

\bibitem[{Kliem {et~al.}(2000)Kliem, Karlicky, \& Benz}]{kliem00}
Kliem, B., Karlicky, M., \& Benz, A.~O. 2000, A\&A, 360, 715

\bibitem[{Kowal {et~al.}(2009)Kowal, Lazarian, Vishniac, \&
  Otmianowska-Mazur}]{kowal09}
Kowal, G., Lazarian, A., Vishniac, E.~T., \& Otmianowska-Mazur, K. 2009, arXiv,
  0903.2052

\bibitem[{Lapenta(2008)}]{lapenta08}
Lapenta, G. 2008, Phys. Rev. Lett., 100, 235001

\bibitem[{Lazarian \& Vishniac(1999)}]{lazarian99}
Lazarian, A., \& Vishniac, E.~T. 1999, ApJ, 517, 700

\bibitem[{Lin {et~al.}(2004)Lin, Raymond, \& van Ballegooijen}]{lin04}
Lin, J., Raymond, J.~C., \& van Ballegooijen, A.~A. 2004, ApJ, 602, 422

\bibitem[{Lin {et~al.}(2008)Lin, Martin, Engvold, Rouppe van~der Voort, \& van
  Noort}]{lin08}
Lin, Y., Martin, S., Engvold, O., Rouppe van~der Voort, L., \& van Noort, M.
  2008, Adv. Space Res., 42, 803

\bibitem[{Loureiro {et~al.}(2007)Loureiro, Schekochihin, \&
  Cowley}]{loureiro07}
Loureiro, N.~F., Schekochihin, A.~A., \& Cowley, S.~C. 2007, Phys. Plasmas, 14,
  100703

\bibitem[{Nitta(2007)}]{nitta07}
Nitta, S.-Y. 2007, ApJ, 663

\bibitem[{Poletto {et~al.}(2004)Poletto, Suess, Bemporad, Schwadron, Elliott,
  Zurbuchen, \& Ko}]{poletto04}
Poletto, G., Suess, S.~T., Bemporad, A., Schwadron, N.~A., Elliott, H.~A.,
  Zurbuchen, T.~H., \& Ko, Y.-K. 2004, ApJ, 613, L173

\bibitem[{{Rappazzo} {et~al.}(2005){Rappazzo}, {Velli}, {Einaudi}, \&
  {Dahlburg}}]{rappazzo05}
{Rappazzo}, A.~F., {Velli}, M., {Einaudi}, G., \& {Dahlburg}, R.~B. 2005, ApJ,
  {\bf 633}, 474

\bibitem[{Scholer(1989)}]{scholer89}
Scholer, M. 1989, J. Geophys. Res., 94, 8805

\bibitem[{Shibata(1996)}]{shibata96}
Shibata, K. 1996, Adv. Space Res., 17, 9

\bibitem[{Shibata {et~al.}(2007)Shibata, Nakamura, Matsumoto, Otsuji, Okamoto,
  Nishizuka, Kawate, Watanabe, Nagata, UeNo, Kitai, Nozawa, Tsuneta, Suematsu,
  Ichimoto, Shimizu, Katsukawa, Tarbell, Berger, Lites, Shine, \&
  Title}]{shibata07}
Shibata, K., {et~al.} 2007, Science, 318, 1591

\bibitem[{Skender \& Lapenta(2009)}]{skender09}
Skender, M., \& Lapenta, G. 2009, ApJ

\bibitem[{Syrovatskii(1981)}]{syrovatskii81}
Syrovatskii, S.~I. 1981, ARA\&A, 19, 163

\bibitem[{Tajima \& Shibata(2002)}]{tajimashibatabook}
Tajima, T., \& Shibata, K. 2002, Plasma astrophysics (Perseus Books Group)

\bibitem[{{Tsuneta}(1996)}]{tsuneta96}
{Tsuneta}, S. 1996, ApJ, {\bf 456}, 840

\bibitem[{Yokoyama \& Shibata(2001)}]{yokoyama01}
Yokoyama, T., \& Shibata, K. 2001, ApJ, 549, 1160

\end{thebibliography}
\end{document}